\documentclass{aa}
\usepackage{lscape,graphicx}
\usepackage{./natbib}
\usepackage{./ctable}
\bibpunct{(}{)}{;}{a}{}{ }

\begin{document}

\title{ Identifying a black hole X-ray transient in M31 with XMM-Newton and Chandra}
\author{ R. Barnard\inst{1}
    \and U. Kolb\inst{1}
    \and J. P. Osborne\inst{2}}

\offprints{R. Barnard, \email{r.barnard@open.ac.uk}}

\institute{ The Department of Physics and Astronomy, The Open University, Walton Hall, Milton Keynes, MK7 6BT, U.K.
     \and The Department of Physics and Astronomy, The University of Leicester, Leicester, LE1 7RH, U.K.}
\date{}

\abstract{Stochastic variability  in two out of four XMM-Newton observations of XMMU\thinspace  J004303+4115 along with its power spectra and X-ray luminosities suggest a low-mass X-ray binary  (LMXB) with a black hole primary. However, Chandra observations resolve the object into two point sources. We use data from 35 Chandra observations  to analyse the contributions of each source, and attribute the variability to \object{CXOM31\thinspace J004303.2+411528} { (known as r2-3)}, which varies in intensity by a factor of $\sim$100 between observations. We assume that the power density spectra  of LMXBs are governed by the luminosity, and that the transition between types of power density spectra occurs at some critical luminosity { in Eddington units,  $l_{\rm c}$}, that applies to all LMXBs. We use results from these XMM-Newton observations and past results from the available literature to estimate this transition  luminosity, and find that all results are consistent with  $l_{\rm c}$ $\sim$0.1  in the 0.3--10 keV band.  CXOM31\thinspace J004303.2+411528  exhibits a low accretion rate power density spectrum at a 0.3--10 keV luminosity of 5.3$\pm0.6$$\times$10$^{37}$ erg s$^{-1}$. Known stellar mass black holes have masses of 4--15 M$_{\odot}$; hence our observations of CXOM31\thinspace J004303.2+411528 are consistent with $l_{\rm c}$ $\sim$0.1 if it has a black hole primary.

 } 

\titlerunning{Black hole transient in M31}
\maketitle

\section{Introduction}
\label{intro}

\begin{table*}[!t]
\centering
\caption{Journal of XMM-Newton observations of the \object{M31} core}\label{journ}
\begin{tabular}{lllllll}
\noalign{\smallskip}
\hline
\noalign{\smallskip}

Observation & Date & MJD&  Exp  & Filter\\
\noalign{\smallskip}
\hline
\noalign{\smallskip}
x1 &  25/06/00 (rev0100)& 51720& 34 ks& Medium \\
x2 & 27/12/00 (rev0193)& 51906 & 13 ks&  Medium\\
x3 & 29/06/01 (rev0285)& 52089& 56 ks &Medium & \\
x4 & 06/01/02 (rev0381)& 52280 & 61 ks&  Thin\\
\noalign{\smallskip}
\hline
\noalign{\smallskip}

\end{tabular}
\end{table*}

The Andromeda Galaxy (M31) is an attractive and important  target for X-ray astronomy, since it is the nearest spiral galaxy \citep[760 kpc, ][]{vdb00}, and its X-ray emission is dominated by point sources. These point sources are thought to be mostly X-ray binaries, along with a few foreground objects, background active galactic nuclei (AGN), and supernova remnants (SNR). The two most recent X-ray observatories, Chandra and XMM-Newton, have finally allowed studies of variability in extra-galactic X-ray sources over time-scales of a few hundred seconds, as well as between successive observations. Such short-term time variability is often characteristic of well-studied phenomena and sometimes allows classification of the objects from X-ray observations alone, in conjunction with their X-ray spectral properties{; for example, thermonuclear X-ray bursts \citep{lvv95}  identify an X-ray source as an { X-ray binary} with a neutron star primary}. To date,  analysis of the XMM-Newton  observations of the core of M31 has resulted in the discovery of   a pulsating supersoft source with a period of 865 seconds \citep{osb01}, the periodic dipping of the X-ray counterpart to the globular cluster Bo\thinspace 158 \citep{tru02}, a persistently bright black hole binary \citep[][ Paper 1]{bok03}, and a Z-source \citep[][ Paper 2]{bko03}. Meanwhile \citet{kaa02} identified variability of three sources in a 47 ks Chandra HRC observation of M31, including the black hole binary later identified in Paper 1.

The XMM-Newton and Chandra missions { complement} each other well; Chandra provides imaging with exceptional spatial resolution, while XMM-Newton is the most sensitive imaging X-ray observatory yet flown. The current work  exemplifies how XMM-Newton and Chandra results can be used together to  get more detailed information than is possible from either data set alone.
XMMU\thinspace  J004303+4115 appears as a point source when observed with XMM-Newton, but is resolved by Chandra into two objects, 6\arcsec~apart. \citet{K02} associate CXOM31\thinspace J004202.9+411523, the southern source,  with { the  globular cluster Bo 146}, and report transient behaviour in CXOM31\thinspace J004303.2+411528, the northern source. {  Following \citet{K02}, we designate the northern source  r2-3 and the southern source r2-4}. We find that in XMM-Newton observations, XMMU\thinspace  J004303+4115 exhibits { power density spectra such as are seen in low accretion rate low-mass X-ray binaries (LMXBs), yet at  0.3--10 keV luminosities of 3--12$\times$10$^{37}$ erg s$^{-1}$ (Sect. 3).}

{
\citet[][ hereafter referred to as vdK94]{vdk94} showed that the power density spectra (PDS) of LMXBs with neutron star or black hole primaries are strikingly similar. At low accretion rates, the PDS of LMXBs have almost identical shapes (approximately broken power laws with the spectral index, $\gamma$, changing from $\sim$0 to $\sim$ 1 at frequencies higher than 0.01--1 Hz) and fractional rms amplitudes of a few times 10\% (vdK94); we shall refer to these as Type A PDS. At higher accretion rates, LMXBs are considerably less variable, with fractional rms amplitudes of only a few percent, and their PDS are described by power laws with $\gamma$ $\sim$1--1.5 (vdK94); we will refer to these as Type B PDS. Furthermore, vdK94 proposed that the transition between Type A and Type B PDS occurs at a critical fraction of the Eddington limit. Following vdK94, we define the critical luminosity fraction, $l_{\rm c}$, as
\begin{equation}
l_{\rm c} = \frac{L_{\rm c}}{L_{\rm Edd}},
\end{equation}
where $L_{\rm c}$ is the luminosity of transition between Type A and Type B PDS, and $L_{\rm Edd}$ is the Eddington luminosity.
}

In Sect. 3 we analyse longterm lightcurves of r2-3 and r2-4 from 35 Chandra observations, paying particular attention to those that were made within 30 days of one of the XMM-Newton observations. If we can associate { Type A} PDS with either source when its luminosity, $L$, exceeds  $L_{\rm c}$ for a neutron star, we can establish that the primary is a black hole. 
 The most likely candidate for such black hole behaviour is r2-3, since most black hole binaries are transients { \citep[e.g.][]{mr03}}, and globular cluster X-ray sources mostly contain { 1.4 M$_{\odot}$} neutron stars \citep[see ][ and references within]{hgl03}.  We present in Sect. 3 evidence that r2-3 contains a black hole primary. 

In Sect. 4 we first establish that r2-3 is located in M31. We then obtain an empirical value for { $l_{\rm c}$}, using results from these XMM-Newton observations of globular cluster sources in M31 and published results from analysis of a Galactic neutron star LMXB and a Galactic black hole LMXB.  {
We then use our value of { $l_{\rm c}$} to calculate $L_{\rm c}$ for  a neutron star with a  mass of 3.1 M$_{\odot}$, the theoretical maximum \citep{kk78}. We assert that r2-3 exhibits a Type A PDS at a luminosity that exceeds this limit and conclude that the primary in r2-3 is a black hole, consistent { with} its transient behaviour. 
}

\section{Observations}
 \label{obs}

\begin{figure}[!b]
\resizebox{\hsize}{!}{\includegraphics{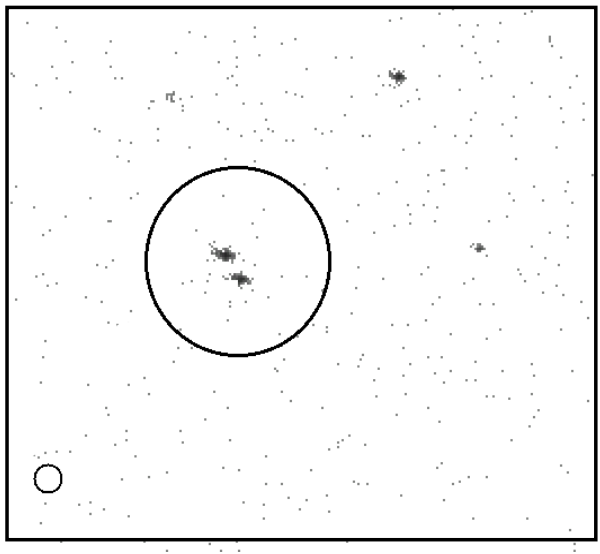}}
\caption{ A detail of the ACIS-I image from the   2000 July 29 Chandra observation of the core of M31; north is up, east is left. The image is log scaled and 2$\arcmin$ across.  The large black circle is centred between { r2-3 (the northern source) and r2-4 (the southern source)} and has a 20$\arcsec$ radius, indicating the extraction region used in the XMM-Newton observations x1--x4. The smaller circle at the bottom left has a radius of 3$\arcsec$; extraction regions of this size were used for r2-3 and r2-4 in observations c1--c35.}\label{obc7}
\end{figure}

\begin{table*}[!t]
\centering
\caption{Chandra  observations of the \object{M31} core. An asterisk denotes observations that were made within 30 days of  an XMM-Newton observation; the relative times of these observations are given in the fourth column.}\label{journc}
\begin{tabular}{llllll}
\noalign{\smallskip}
\hline
\noalign{\smallskip}

Obs. & Date  (Obs ID)& MJD & Mode&  Exposure & Separation \\
\noalign{\smallskip}
\hline
\noalign{\smallskip}
c1 & 13/10/99  (303)&51464& ACIS-I& 12 ks&  \\
c2 & 30/11/99 (267) & 51512 &HRC-I& 1.3 ks\\
c3 & 11/12/99 (305)& 51523& ACIS-I & 4.2 ks \\
c4 & 23/12/99 (268) & 51535& HRC-I & 5.2 ks\\
c5 & 27/12/99  (306)& 51539& ACIS-I & 4.2 ks \\
c6 & 19/01/00 (269) & 51562 & HRC-I & 1.2 ks\\
c7 & 29/01/00 (307)& 51572& ACIS-I& 4.2 ks\\
c8 & 13/02/00 (270) & 51587& HRC-I& 1.5 ks\\
c9 & 16/02/00 (308) & 51590&ACIS-I& 4.1 ks\\
c10 & 08/03/00 (271) & 51611& HRC-I& 2.5 ks\\
c11$^*$ & 26/05/00 (272) & 51690& HRC-I & 1.2 ks & $-$30 days\\
c12$^*$ & 01/06/00 (309)&51696& ACIS-S & 5.2 ks & $-$24 days\\
c13$^*$ & 21/06/00 (273) & 51716& HRC-I & 1.2 ks & $-$4 days\\
c14$^*$  & 02/07/00 (310)&51727& ACIS-S  &5.1 ks & +7 days \\
c15 & 29/07/00 (311) & 51754& ACIS-I& 5.0 ks \\
c16 & 18/08/00 (275) & 51774 & HRC-I & 1.2 ks\\
c17 & 27/08/00 (312) & 51783& ACIS-I& 4.7 ks\\
c18 & 11/09/00 (276) & 51798 & HRC-I & 1.2 ks\\
c19 & 12/10/00 (277) & 51829 & HRC-I & 1.2 ks \\
c20 & 17/11/00 (278) & 51865 & HRC-I & 1.2 ks \\
c21$^*$ & 31/11/00 (1912) & 51879 & HRC-I & 47 ks & $-$27 days\\
c22$^*$ & 13/12/00 (1581) & 51891& ACIS-I& 4.5 ks& $-$15 days\\
c23$^*$ & 13/01/01 (1854) & 51922& ACIS-S& 4.8 ks& +16 days\\
c24 & 01/02/01 (1569) & 51941 & HRC-I & 1.2 ks\\
c25 & 18/02/01 (1582)& 51958& ACIS-I& 4.4 ks\\
c26$^*$ &10/06/01 (1570) & 52070 & HRC-I & 1.2 ks & $-$19 days\\ 
c27$^*$ & 10/06/01 (1583)& 52070& ACIS-I & 5.0 ks& $-$19 days\\
c28 & 31/08/01 (1577) & 52152& ACIS-I& 5.0 ks \\
c29 & 09/11/01 (1585) & 52222& ACIS-I& 5.0 ks \\
c30 & 19/11/01 (2904) & 52232& HRC-I& 1.2 ks \\
c31$^*$ &08/01/02 (2897) & 52282& ACIS-I& 5.0 ks & +1.5 days\\
c32$^*$ & 16/01/02 (2905) & 52290& HRC-I & 1.1 ks & +10 days\\
c33 & 02/06/02 (2906) & 52427 & HRC-I & 1.2 ks \\
c34 & 02/06/02 (2898) & 52427 & ACIS-I & 5.0 ks \\
c35 & 11/08/02 (4360) & 52497 & ACIS-I& 5.0 ks \\
   \noalign{\smallskip}
\hline
\noalign{\smallskip}
\end{tabular}
\end{table*}

Four XMM-Newton  \citep{jan01} observations were made of the core of M31; details of the  observations are given in Table~\ref{journ}. For each observation, 0.3--10 keV lightcurves were extracted from source and background regions for each of the three EPIC instruments: one PN camera \citep{stru01} and two MOS cameras \citep{turn01}. The source region was circular with a radius of 20$\arcsec$ and centred on XMMU\thinspace J004303+4115 while the background region was an equivalent source-free region on the same CCD and at a similar offset from the optical axis. The lightcurves were accumulated with 2.6 second binning. Background-subtracted, summed EPIC lightcurves,  were then produced following the procedures laid out in Paper 1. PN energy spectra from the source and background regions in the 0.3--10 keV energy band were also extracted for observations x1--x4, as described in Paper 1. Strong { solar} flaring contaminated parts of observations x3, restricting the good data to $\sim$25 ks.

In order to assess the contributions of the two sources seen with Chandra to the properties of XMMU\thinspace  J004303+4115, we analysed 35 Chandra observations; 15  observations used the ACIS-I configuration, 3 used ACIS-S and 17 were made with the  HRC-I. Of these, 11 observations occurred at similar times to the XMM-Newton observations.  From the 35 observations, we produced long-term lightcurves for r2-3 and r2-4 in order to study the variability of these objects. In particular, we examined their contributions to the observations close to the XMM-Newton observations, to associate the variability and contributions to the X-ray luminosity with a particular source.

\begin{figure}[!t]
\resizebox{\hsize}{!}{\includegraphics{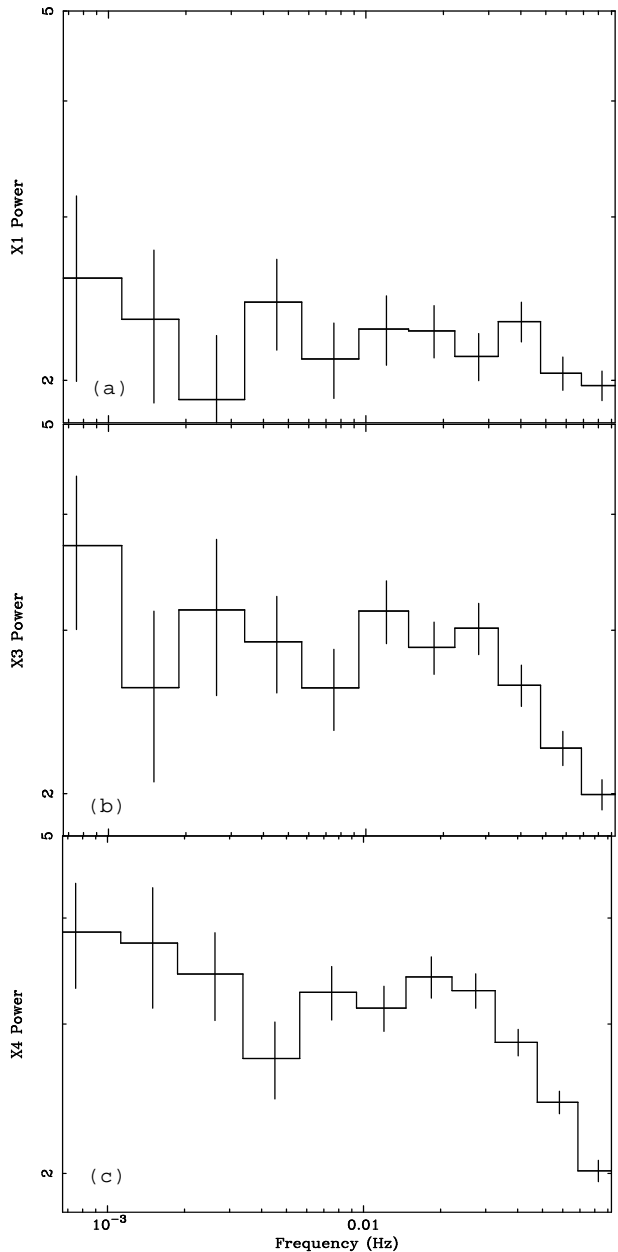}}
\caption{Power density spectra from the 0.3--10 keV lightcurves of XMMU\thinspace  J004303+4115 from observations x1 (panel a), x3 (panel b) and x4 (panel c). We see that there is more low-frequency power, and a steeper drop at high frequencies, in { x3 and  x4 than in x1}. The PDS are Leahy normalised, so the Poisson noise has a power of 2. We believe the PDS of observation x1 is due to r2-4 alone (Sect.~\ref{longlc}).}\label{po}
\end{figure}

 The date, configuration mode and exposure for each Chandra observation are listed in Table~\ref{journc}; { particular attention was paid to those Chandra observations within 30 days of an XMM-Newton observation: this interval was chosen to maximise the data available on r2-3 and r2-4 at
similar epochs to x1-x4, and is not physically motivated}.
Chandra lightcurves  were extracted in the 0.3--7.0 keV energy band  from circular extraction regions with 3$\arcsec$ radii around r2-3 and r2-4 for each  observation, with 50 second binning. To do this, a binned image was created using {\sc dmcopy}; ACIS and HRC images were binned by factors of 2 and 10 respectively. Events files were then created in the 0.3--7 keV band  for r2-3 and r2-4 with {\sc dmcopy}, filtering with the relevant extraction region. Finally, lightcurves were created for each source using {\sc lightcurve}.
Lightcurves from nearby, source-free regions of the same  size yielded $\leq$ 1 photon per observation, hence the background contribution was negligible. Long-term lightcurves were then constructed for r2-3 and r2-4, representing each observation with a single bin.  A detail of the ACIS-I image from observation c15 is presented in Fig.~\ref{obc7}; r2-3 and r2-4 are shown, and { a circle with a 20$\arcsec$ radius}  is centred between them, defining the source extraction region for the XMM-Newton observations. A 3$\arcsec$ circle is shown in the bottom left corner to indicate the size of the extraction regions used in the Chandra observations.

X-ray spectra were extracted from ACIS-I and ACIS-S observations of r2-3 and r2-4 using {\sc psextract}, which also produced corresponding response matrices and ancillary response files. These spectra were grouped to have a minimum of 15 counts per bin, or 10 counts per bin if the total number of counts $\leq$100. No spectra were obtained from HRC-I because it is not designed for spectral work.

\section{Results}
\label{res}

\subsection{XMM--Newton Lightcurves and power density spectra}
Lightcurves with 200 second bins  of XMMU\thinspace  J004303+4115 from observations x3 and x4 were found to be highly variable; best $\chi^2$ fit lines of constant intensity to the data were rejected at confidence levels of  0.002\% and $\ll$5$\times$10$^{-6}$\% respectively.  However, the similarly binned lightcurves from x1 and x2 were not significantly variable over this time scale.

Power density spectra  were obtained from the 0.3--10 keV lightcurves of observations x1, x3 and x4, with (10.4 s)$^{-1}$ resolution and 512 frequency bins, as described in Paper 1.
The resulting PDS are presented in Fig.~\ref{po}; observation x2 was too short to yield useful PDS data. We fitted the PDS with single power laws, and while the PDS of observation x1 is well fitted ($\chi^2$/dof = 7.7/9), fits to the PDS of  x3 ($\chi^2$/dof = 26/9) and x4 ($\chi^2$/dof = 64/9) are rejected at levels of 0.2\% and $\ll$5$\times$10$^{-6}$\% respectively. 
We see that a { Type B} PDS is observed in x1, where the  combined EPIC intensity  of XMMU\thinspace J004303+4115 is 0.16 count s$^{-1}$ in the 0.3--10 keV band and yet a { Type A} PDS is observed in x4, where its intensity is 0.37 count s$^{-1}$, i.e. over twice as bright as in x1. We also see a { Type A} PDS in observation x3, where the intensity is 0.15 count s$^{-1}$, very similar to the count rate in x1; hence we know that the lack of a broken PDS in x1 is not due to a lack of counts. Since we know that XMMU\thinspace J004303+4115 is in fact composed of two X-ray sources, we can only understand these observations if the component responsible for the { Type A} PDS is absent in x1 but present in x3 and x4.

\subsection{Energy Spectra}
\label{enspec}
PN spectra of XMMU\thinspace  J004303+4115 from observations x1--x4 were fitted with simple emission models to estimate the 0.3--10 keV flux, and hence the combined luminosity of the two objects. The model consisted  of a single power law suffering line-of-sight absorption. Although good fits were obtained, this was not a physically motivated model, it was simply used to estimate the luminosity of the source in the X-ray band. The luminosity of XMMU\thinspace  J004303+4115 ranged over 3.0--12.0$\times$10$^{37}$ erg s$^{-1}$. Details of the fits are given in Table~\ref{xspecfits}. The intensities of XMMU\thinspace  J004303+4115 in observations x1--x4 were converted to Chandra ACIS-I count rates using {\sc pimms} and the best model parameters.

\begin{figure}[!t]
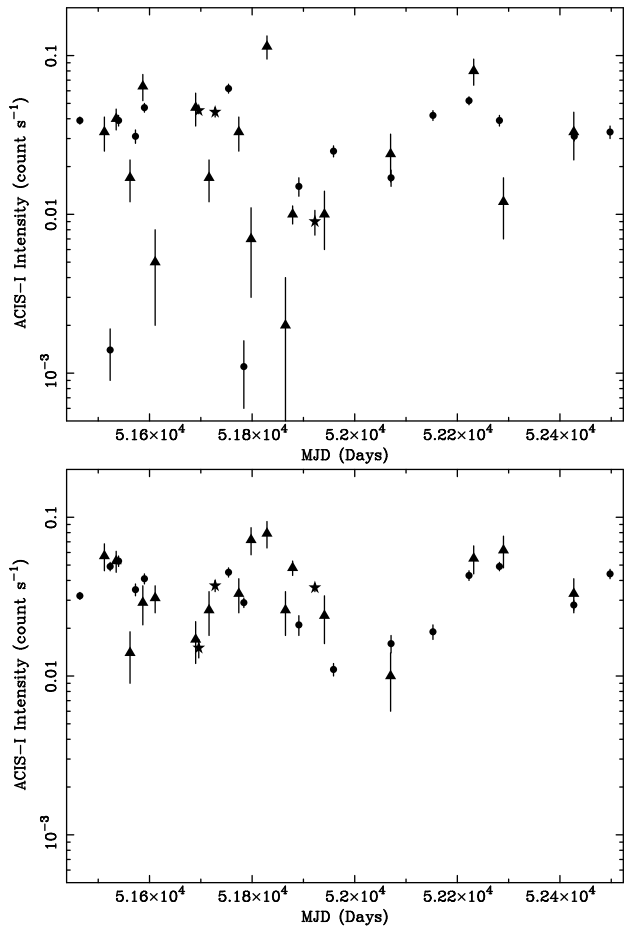

\resizebox{\hsize}{!}{\includegraphics[angle=270]{aa20030661_fig3a.ps}}
\resizebox{\hsize}{!}{\includegraphics[angle=270]{aa20030661_fig3b.ps}}
\caption{Long term lightcurve from observations c1--c35 of r2-3 (top)  and r2-4 (bottom), converted to ACIS-I intensity. The y-axes are log-scaled. Stars represent ACIS-S observations, circles represent ACIS-I observations and triangles represent HRC-I observations.}\label{clc}
\end{figure}

 We then analysed the energy spectra from the Chandra ACIS-I and ACIS-S observations. The line-of-sight absorption was not well constrained  and was fixed at 1.0$\times$10$^{21}$ atom cm$^{-2}$, to agree with XMM-Newton values. The spectral index of r2-3 varied over the range 1.6--2.6, while the spectral index of r2-4 ranged over 1.3--2.3. The best fit spectral index, $\chi^2$/dof and 0.3-7 keV fluxes are provided for r2-3 and r2-4 in Table~\ref{cspecfit}.  There was no significant correlation between spectral index and intensity within the limits of the data, so the mean values were used in converting HRC-I intensities into ACIS-I intensities: spectral indices of 1.78$\pm$0.10 and 1.73$\pm$0.08 were used for r2-3 and r2-4 respectively.

\subsection{Longterm lightcurves}
\label{longlc}
The long-term  Chandra lightcurves of r2-3 and r2-4 are presented in Fig.~\ref{clc}.  The 0.3--7 keV intensities of the ACIS-S observations were converted to 0.3--7 keV ACIS-I intensities with {\sc pimms}, using the best fit absorbed power law models. { Longterm lightcurves of r2-3 and r2-4 are also presented by \citet{will03}, scaled to the luminosity}. The 1$\sigma$ uncertainties on converting  the measured XMM-{ Newton}, HRC-I and ACIS-S intensities to ACIS-I intensities were obtained as follows. {\sc pimms} was used to obtain the uncertainty in the ACIS-I intensity due to the 1$\sigma$ uncertainty in the spectral index of the original fits. These were added in quadrature to the statistical count rate uncertaintes.
We see that the intensity of r2-3 varies by a factor of $\sim$100, from 0.0011 to 0.114 count s$^{-1}$, while r2-4 varies more modestly, from 0.010 to 0.079 count s$^{-1}$. 

{ The longterm lightcurve of r2-3 is quite unlike the lightcurves of classical black hole LMXBs, which are characterised by short outbursts of activity where the intensity decays exponentially over several weeks, separated by years of quiescence \citep{mr03}. Of the Galactic black hole LMXBs, it most resembles GX\thinspace 339$-$4 \citep[dynamically confirmed as a black hole by][]{hyn03}. \citet{zdz04} present long-term lightcurves of GX\thinspace 339$-$4 spanning $\sim$20 years from the Ginga, RXTE and CGRO satellites; around 15 outbursts were observed, varying dramatically in brightness and duration, and in the separation between  outbursts. The behaviour of  r2-3 (Fig.~\ref{clc}) is similar to that exhibited by GX\thinspace 339$-$4 in the interval MJD $\sim$50200--51000, where GX\thinspace 339$-$4 appears to be mostly active, with short intensity dips \citep{zdz04}. 

The longterm lightcurve of r2-4, on the other hand,  shows modulation on a $\sim$400 day time-scale. It is similar to the longterm RXTE ASM lightcurve of \object{4U\thinspace 1820$-$30}, an ultracompact neutron star LMXB in a Galactic globular cluster \citep{stel87}. This source exhibits brief low intensity states superposed onto cyclical variability on a $\sim$170 day period featuring a sharp rise and shallow decay \citep{sim03}. 

}

{
\ctable[
caption ={ Best fit parameters for the 0.3--10 keV PN spectra of XMMU\thinspace  J004303+4115\tmark[a]. Numbers in parentheses represent 90\% confidence uncertainties on the last digit. },
label = {xspecfits},
pos=b,
]{lllll}{
\tnote[a]{ an absorbed power law model was used, with line-of-sight absorption
 $N_{\rm H}$ (in units of 10$^{21}$ atom cm$^{-2}$) and spectral index $\alpha$}
\tnote[b]{ 0.3--10 keV luminosity / 10$^{37}$ erg s$^{-1}$}
}
{                                                \FL
Obs. & $N_{\rm H}$ &$\alpha$  & $L$\tmark[b] & $ \chi^2$/dof\ML
x1 & 0.6(2) &  2.15(15) & 3.0(2) &37/34\NN
x2 & 1.0(2) & 1.91(15)  & 4.4(3) & 36/34\NN
x3 & 1.1(2) & 1.98(15) & 3.3(2) & 16/18\NN
x4 & 1.4(2)& 1.89(15) & 12.0(7) & 17/34\LL
}
}

\begin{figure}[!t]
\resizebox{\hsize}{!}{\includegraphics[angle=270]{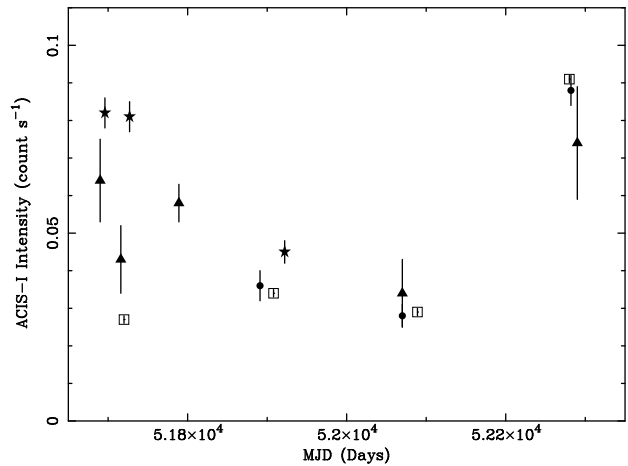}}
\caption{ A comparison of XMM-Newton observations of XMMU\thinspace  J004303+4115 ({ open} squares) with neighbouring  Chandra ACIS-I, ACIS-S  and HRC-I observations. The XMM-Newton intensities were converted to ACIS-I count rates using {\sc pimms}; for conversion we assumed a power law spectral model, using the best fit parameters listed in Table~\ref{xspecfits}.}\label{xcomp}
\end{figure}

{ The ACIS-I equivalent intensities of r2-3 + r2-4 in x1-x4 and the nearby Chandra observations are presented in Fig~\ref{xcomp}.}
 Table~\ref{xns} lists the intensities of XMMU\thinspace J004303+4115 in observations x1--x4  with the individual intensities of r2-3 and r2-4 in neighbouring Chandra observations; {  the fractional contribution of r2-3 to the total intensity is also given.}
Observations  x3 and x4 have very similar intensities to the neighbouring Chandra observations (within 0.5$\sigma$), so the fractional contributions of r2-3 and r2-4 to these XMM-Newton observations were { assumed to be the same as in the } nearest Chandra observation. { However, the intensity in x1 is well below the intensities of any neighbouring Chandra observation; the intensity dropped by 67\% (11$\sigma$) in 24 days then increased by 200\% (12$\sigma$) in 7 days.
  Instead, the intensity { and X-ray spectrum} of x1 are consistent with that of r2-4 in c13, suggesting that r2-3 has all but disappeared, as in observations c3,  c17 and c20 { where the significance of detecting r2-3 is $<$3$\sigma$}.

 This interpretation is supported by the fact that r2-3 { contributed} $\sim$70\% of the intensity in c11 and c12, but only 35$\pm$13\% in c13 (Table~\ref{xns}), rising to 55$\pm$5\% in c14.  The spectral index, $\alpha$, of r2-3 is 1.7$\pm$0.2 in c12 and c14, while for r2-4, $\alpha$ = 2.3$\pm$0.4 in c12 and 1.5$\pm$0.2 in c14. In x1, $\alpha$ = 2.15$\pm$0.15, which is considerably softer than in x2--x4 and consistent with that of r2-4 alone. 
} We cannot be sure how the two sources contribute to XMMU\thinspace J004303+4115 in observation x1, but given the observed trends in observations c12--c14 the most likely explanation of the 11$\sigma$ drop in intensity between c12 and x1 is the fading to invisibility of r2-3.

\ctable[
caption={Best fit models to the 0.3--7 keV ACIS-I and ACIS-S  spectra of r2-3 and r2-4\tmark[a]. Numbers in parentheses represent 90\% confidence uncertainties on the last digit.},
label={cspecfit},
star,
pos=t,
]
{lllllll}{
\tnote[a]{ an absorbed power law model was used, with line-of-sight absorption $N_{\rm H}$ = 1.0$\times$10$^{21}$ atom cm$^{-2}$ and spectral index $\alpha$}
\tnote[b]{ 0.3--10 keV luminosity / 10$^{37}$ erg s$^{-1}$}
}
{\FL
Observation &$\alpha_{r2-3}$ & $L_{\rm r2-3}$\tmark[b] & $\chi^2$/dof &$\alpha_{\rm r2-4}$  & $L_{\rm r2-4}$\tmark[b] & $\chi^2$/dof\ML
c1 & 2.1(2) & 3.3 & 11/18 & 1.7(2) & 2.4  & 13/14 \NN
c3 & $\dots$ & $\dots$ & $\dots$ & 1.4(3) &  3.9 & 16/10\NN
c5 & 1.9(3) & 2.5 & 16/12 & 1.5(3) &  4.1 & 20/12\NN
c7 & 2.1(4) & 2.2 & 6/9 & 1.8(3) & 2.5 & 6/11\NN
c9 & 1.7(3) & 3.4 & 15/15 & 1.5(4) & 3.4 & 12/6\NN
c12 & 1.7(2) & 3.0 & 20/11 & 2.3(4) & 0.9 & 6/7\NN
c14 & 1.7(2) & 4.0 & 11/11 & 1.5(2) & 3.0 & 10/12\NN
c15 & 1.6(2) & 4.5 & 11/16 &  1.8(2) & 3.0 & 12/11\NN
c17 & $\dots$ & $\dots$ & $\dots$ & 1.6(2) & 2.0 & 11/10\NN
c22 & 2.2(9) & 2.0 & 4/3 & 1.9(5) & 1.8 & 3/6 \NN
c23 & 2.6(7) & 0.7 & 1/3 & 1.6(2) & 4.0 & 4/11\NN
c28 & 1.8(3) & 3.8 & 8/10 & 2.2(6) & 1.9 & 10/6\NN
c29 & 1.5(3) & 4.4 & 6/9 & 1.4(2) & 3.5 & 11/11\NN
c31 & 2.1(3) & 3.0 & 11/9 & 1.3(2) & 3.7 & 13/9\NN
c34 & 1.9(3) & 2.3 & 6/7 & 2.3(4) & 2.0 & 8.5/10\NN
c35 & 2.0(4) & 2.7 & 3/7 &  1.7(3) &  3.5 & 5/7\LL
}

\ctable[
caption={ List of intensities of XMMU\thinspace J004303+4115 in observations x1--x4 and the individual intensities of r2-3 and r2-4 in neighboring observations. Numbers in parenthesis indicate 1 $\sigma$ uncertainties on the last digits.},
label={xns},
star,
]
{llllllll}
{
\tnote[a]{ Equivalent 0.3--7 keV ACIS-I intensity / count s$^{-1}$}
\tnote[b]{ $f_{\rm r2-3}$ = $I_{\rm r2-3}$ / ($I_{\rm r2-3} + I_{\rm r2-4})$}
\tnote[c]{ $I_{\rm X}$ = $I_{\rm r2-3}$ + $I_{\rm r2-4}$ for observations x1--x4}
}
{\FL
Observation & MJD  &$I_{\rm r2-3}$\tmark[a]&  $I_{\rm r2-4}$\tmark[a] & $f_{\rm r2-3}$\tmark[b] & $I_{\rm X}$\tmark[c] \ML
c11  & 51690 & 0.047(10) & 0.017(5) & 0.7(2) &  $\dots$\NN
c12  & 51696 & 0.061(3) & 0.021(2) & 0.74(5) & $\dots$ \NN
c13  & 51716 & 0.017(5) & 0.026(8) & 0.35(13) &$\dots$\NN
x1 & 51720 & $\dots$ & $\dots$ & $\dots$ & 0.027(1)\NN
c14  & 51727 & 0.044(3) & 0.037(3) & 0.55(5) & $\dots$\NN
\addlinespace[.2cm]
c21  & 51879 & 0.010(1) & 0.048(5) & 0.17(2) & $\dots$\NN
c22  & 51891 & 0.015(2) & 0.021(3) & 0.42(7) & $\dots$\NN
x2 & 51908 & $\dots$ & $\dots$ & $\dots$ & 0.034(1)\NN
c23  & 51922 & 0.009(2) & 0.036(2) & 0.20(5)& $\dots$\NN
\addlinespace[.2cm]
c26 & 52070 & 0.024(8) & 0.010(4)& 0.7(3) &$\dots$\\
c27 & 52070 & 0.017(2) & 0.016(2)& 0.52(8) & $\dots$\\
x3 & 52089 & $\dots$ & $\dots$ & $\dots$ & 0.029(1)\\
\addlinespace[.2cm]
x4 & 52280 & $\dots$ & $\dots$ & $\dots$ & 0.091(1)\\ 
c31 & 52282 & 0.039(3) & 0.049(3) & 0.44(4)& $\dots$\\
c32 & 52290 & 0.012(5) & 0.062(14) & 0.16(7) & $\dots$\LL
}

\subsection{Overall behaviour}
The projected 0.3--10 keV luminosities of r2-3 and r2-4 in observations { x1, x3 and  x}4 are given in Table~\ref{lns}; also, { the  PDS type} is indicated. 
In observations x3 and x4, where we know that r2-3 contributes, { Type A PDS are seen}.  We argue that r2-3 is absent in observation x1, where  a { Type B PDS is observed}. The { Type A} PDS observed in x3 could come from either source, since the projected luminosity of r2-4 is  { half that} in observation x1, where the { Type B} PDS is seen. However, r2-4 could not exhibit a { Type B} PDS at 3$\times$10$^{37}$ erg s$^{-1}$ and a { Type A} PDS at $\sim$7$\times$10$^{37}$ erg s$^{-1}$, its luminosity in x4. We conclude that the { Type A} contribution to the PDS of XMMU\thinspace J004303+4115 in x3 and x4 was made by r2-3, { even at a 0.3--10 keV luminosity of 5.3$\pm$0.6$\times$10$^{37}$ erg s$^{-1}$} in x4. We argue in Sect. 4 that a LMXB exhibiting a { Type A} PDS at such a high luminosity is likely to contain a black hole.

\section{Discussion}

In diagnosing the nature of   r2-3 we first need to be certain that it is indeed within the M31 galaxy. We applied a column density of 1.0$\times$10$^{21}$ cm$^{-2}$  and a spectral index of 1.78 to the HRC-I observations that yielded the maximum and minimum ACIS-I { equivalent} count rates of r2-3 and obtained a flux range of  0.012--1.4$\times$10$^{-12}$ erg cm$^{-2}$ s$^{-1}$ in the 0.3--10 keV band using {\sc pimms}. This corresponds to a luminosity range of 0.08--9.5$\times$10$^{37}$ erg s$^{-1}$ for a location in M31. If it was local, it would be most likely to be within 1 kpc, i.e. $<$3 times the  scale-height of the Galactic disc, since M31 is 21.6$\degr$ out of the Galactic plane and 120$\degr$ from the Galactic Centre. In this situation, the luminosity would be $>$6 orders of magnitude smaller, and hence up to 2 orders of magnitude fainter than the faintest known persistent Galactic LMXB \citep{wil03}.
It would also be up to a factor of $\sim$3 fainter than the faintest black hole X-ray transient in quiescence \citep[][ and references within]{tom03}. It is likely that if r2-3 { were} local, it would have an optical counterpart; known absolute V magnitudes (M$_{\rm V}$) of persistent Galactic LMXBs range from  $-$2.5 to 5.6 \citep{vm95}, and the spectral types of the secondary stars in most black hole X-ray transients have been obtained in quiescence \citep{cc03}, with { M$_{\rm V}$} $\sim$0--9.7. Hence, { if r2-3 were local, we would expect to see an optical counterpart with  m$_{\rm V}$ $\la$ 19}. However, the nearest optical source in the HST catalogue of \citet{hai94} is 79$\arcsec$ away, with a { m$_{\rm v}$ =  20.6}; hence any optical counterpart to r2-3 would have to be fainter than this. Thus r2-3 is unlikely to be local.
In addition, its PDS  rules out the possibility that it is a background AGN, as they exhibit spectral breaks at 10$^{-6}$--10$^{-5}$ Hz \citep{utt02} rather than at the $\sim$0.03 Hz seen here. We therefore conclude that r2-3 is located in M31.

 By showing that we observe a { Type A} PDS from r2-3 at a luminosity that { is} too high for a neutron star LMXB we can classify the primary as a black hole. To do this, we must obtain a value for { $l_{\rm c}$, and assume that this applies to all LMXBs. Then we must show that r2-3 exhibits a Type A PDS at a luminosity greater than $L_{\rm c}$ for any neutron star.}

In the first instance, we looked at seven X-ray sources associated with globular clusters in { x4}, since they are likely to contain 1.4 M$_{\odot}$ neutron stars. None of  { them exhibited Type A PDS; the  luminosity range was } $\sim$2--10$\times$10$^{37}$ erg s$^{-1}$ (Paper 1). Also,  a { Type B} PDS was exhibited by XMMU\thinspace J004303+4115 
in observation x1 at 3$\times$10$^{37}$ erg s$^{-1}$ (which we associate with r2-4 alone, see Sect. 3).  These results suggest that {  $L_{\rm c}$}$\la$ 2$\times$10$^{37}$ erg s$^{-1}$; hence {  $l_{\rm c}$ $\la$ 0.1, assuming hydrogen accretion onto a 1.4 M$_{\odot}$ neutron star.}

We obtained a vital clue to { $l_{\rm c}$} from the Galactic neutron star \object{LMXB 4U\thinspace 1705$-$44}.
 \citet{lang89} analysed data from four EXOSAT observations of  4U\thinspace 1705$-$44; they found that it exhibited a { Type A PDS in the faintest observation but a Type B PDS in the next faintest}. In their previous analysis of the observations \citep{lang87}, they obtained 1--11 keV fluxes of 1.3$\times$10$^{-9}$ and 1.8$\times$10$^{-9}$ erg { cm$^{-2}$} s$^{-1}$ for these two observations. Hence an accurate distance to 4U\thinspace 1705$-$44 would yield a tight constraint on { $l_{\rm c}$}. \citet{cs97} estimate a distance of 11 kpc, using the most luminous X-ray burst as a standard candle; this constitutes an upper limit to the distance \citep[see e.g.][ and references within]{kul03}. \citet{cor03} also estimate the distance to 4U\thinspace 1705$-$44 using bursts, but give a distance of 8.9 kpc with an assumed uncertainty of 30\%. Hence { $l_{\rm c}$} = 0.08$^{+0.08}_{-0.05}$.

{  Similarly, \object{GS\thinspace 2023+338} (\object{V404 Cygni}) is a Galactic black hole LMXB; the most likely mass for the primary is 12 M$_{\odot}$ \citep{shab94}. It was discovered with Ginga during an outburst in 1989 \citep{mak89} and exhibited Type A PDS at 2--37 keV luminosities $>$3$\times$10$^{38}$ erg s$^{-1}$ \citep[][ and references within]{miy92,oost97}. The X-ray spectrum for V404 Cygni was described by a power law with $\alpha$ = 1.0--1.4 \citep{miy92}, hence  $l_{\rm c}$ $\ga$ 0.06 in the 0.3--10 keV band.
}

These three sets of results  are all consistent with { $l_{\rm c}$ $\sim$0.10 in the 0.3--10 keV band; this supports the idea of a constant $l_{\rm c}$ proposed by vdK94}.

  { Now, r2-3 appears to exhibit a Type A  PDS at a 0.3--10 keV luminosity of  5.3$\pm$0.6$\times$10$^{37}$ erg s$^{-1}$;  this is a factor of $\sim$3 higher than $L_{\rm c}$ for a 1.4 M$_{\odot}$ neutron star.  Indeed, for the maximum mass of a neutron star \citep[i.e. 3.1 M$_{\odot}$,][]{kk78}, $L_{\rm c}$ $\sim$4$\times$10$^{37}$ erg s$^{-1}$. {  Known stellar-mass black holes have masses over the range 4--15 M$_{\odot}$, hence our results from r2-3 are consistent with $l_{\rm c}$ = 0.1 if the primary is a black hole. }

}

\begin{acknowledgements}
{ The authors would like to thank the anonymous referee for suggestions that led to significant improvements to the paper}. This work is supported by PPARC.
\end{acknowledgements}

\ctable[
caption={Projected 0.3--10 keV luminosities of r2-3 and r2-4 ($L_{\rm r2-3}$ and $L_{\rm r2-4}$ respectively) in terms of 10$^{37}$ erg s$^{-1}$. The  PDS type is also listed.},
label={lns},
pos=h,
]
{llll}
{
}
{\FL
Observation  &$L_{\rm r2-3}$&  $L_{\rm r2-4}$ & Type \ML
x1 & 0 & 3 &  B\NN
x3 & 1.7 & 1.6 & A\NN
x4 & 5.3 & 6.7 &  A\LL 
}

\bibliographystyle{./bibtex/aa}
\bibliography{m31}

\begin{thebibliography}{35}
\expandafter\ifx\csname natexlab\endcsname\relax\def\natexlab#1{#1}\fi

\bibitem[{{Barnard} {et~al.}(2003{\natexlab{a}}){Barnard}, {Osborne}, {Kolb},
  \& {Borozdin}}]{bok03}
{Barnard}, R., {Osborne}, J.~P., {Kolb}, U., \& {Borozdin}, K.~N.
  2003{\natexlab{a}}, \aap, 405, 505
\bibitem[{{Barnard} {et~al.}(2003{\natexlab{b}}){Barnard}, {Kolb}, \&
  {Osborne}}]{bko03}
{Barnard}, R., {Kolb}, U., \& {Osborne}, J.~P. 2003{\natexlab{b}}, \aap, 411,
  553



\bibitem[{{Charles} \& {Coe}(2003)}]{cc03}
{Charles}, P.~A. \& {Coe}, M.~J. 2003, {Compact Stellar Sources} ({Cambridge
  University Press}), in press, astro--ph/0308020

\bibitem[{{Christian} \& {Swank}(1997)}]{cs97}
{Christian}, D.~J. \& {Swank}, J.~H. 1997, ApJ Si, 109, 177+

\bibitem[{{Cornelisse} {et~al.}(2003){Cornelisse}, {in't Zand}, {Verbunt},
  {Kuulkers}, {Heise}, {den Hartog}, {Cocchi}, {Natalucci}, {Bazzano}, \&
  {Ubertini}}]{cor03}
{Cornelisse}, R., {in't Zand}, J.~J.~M., {Verbunt}, F., {et~al.} 2003, \aap,
  405, 1033

\bibitem[{{Haiman} {et~al.}(1994){Haiman}, {Magnier}, {Lewin}, {Lester}, {van
  Paradijs}, {Hasinger}, {Pietsch}, {Supper}, \& {Truemper}}]{hai94}
{Haiman}, Z., {Magnier}, E., {Lewin}, W.~H.~G., {et~al.} 1994, \aap, 286, 725

\bibitem[{{Heinke} {et~al.}(2003){Heinke}, {Grindlay}, {Lugger}, {Cohn},
  {Edmonds}, {Lloyd}, \& {Cool}}]{hgl03}
{Heinke}, C.~O., {Grindlay}, J.~E., {Lugger}, P.~M., {et~al.} 2003, \apj, 598,
  501

\bibitem[{{Hynes} {et~al.}(2003){Hynes}, {Steeghs}, {Casares}, {Charles}, \&
  {O'Brien}}]{hyn03}
{Hynes}, R.~I., {Steeghs}, D., {Casares}, J., {Charles}, P.~A., \& {O'Brien},
  K. 2003, \apjl, 583, L95

\bibitem[{{Jansen} {et~al.}(2001){Jansen}, {Lumb}, {Altieri}, {Clavel}, {Ehle},
  {Erd}, {Gabriel}, {Guainazzi}, {Gondoin}, {Much}, {Munoz}, {Santos},
  {Schartel}, {Texier}, \& {Vacanti}}]{jan01}
{Jansen}, F., {Lumb}, D., {Altieri}, B., {et~al.} 2001, \aap, 365, L1

\bibitem[{{Kaaret}(2002)}]{kaa02}
{Kaaret}, P. 2002, \apj, 578, 114

\bibitem[{{Kong} {et~al.}(2002){Kong}, {Garcia}, {Primini}, {Murray},
  {DiStefano}, \& {McClintock}}]{K02}
{Kong}, A.~H.~K., {Garcia}, M.~R., {Primini}, F.~A., {et~al.} 2002, ApJ, 577,
  738

\bibitem[{{Krishan} \& {Kumar}(1978)}]{kk78}
{Krishan}, V. \& {Kumar}, N. 1978, \apss, 57, 241

\bibitem[{{Kuulkers} {et~al.}(2003){Kuulkers}, {den Hartog}, {in't Zand},
  {Verbunt}, {Harris}, \& {Cocchi}}]{kul03}
{Kuulkers}, E., {den Hartog}, P.~R., {in't Zand}, J.~J.~M., {et~al.} 2003,
  \aap, 399, 663

\bibitem[{{Langmeier} {et~al.}(1989){Langmeier}, {Hasinger}, \&
  {Truemper}}]{lang89}
{Langmeier}, A., {Hasinger}, G., \& {Truemper}, J. 1989, \apjl, 340, L21

\bibitem[{{Langmeier} {et~al.}(1987){Langmeier}, {Sztajno}, {Hasinger},
  {Truemper}, \& {Gottwald}}]{lang87}
{Langmeier}, A., {Sztajno}, M., {Hasinger}, G., {Truemper}, J., \& {Gottwald},
  M. 1987, \apj, 323, 288

\bibitem[{{Lewin} {et~al.}(1995){Lewin}, {van Paradijs}, \& {van den
  Heuvel}}]{lvv95}
{Lewin}, W.~H.~G., {van Paradijs}, J., \& {van den Heuvel}, E.~p.~J. 1995,
  {X-ray Binaries} ({Cambridge University Press})

\bibitem[{{Makino}(1989)}]{mak89}
{Makino}, F. 1989, in International Astronomical Union Circular, 1--+

\bibitem[{{McClintock} \& {Remillard}(2003)}]{mr03}
{McClintock}, J.~E. \& {Remillard}, R.~A. 2003, astro-ph/0306213

\bibitem[{{Miyamoto} {et~al.}(1992){Miyamoto}, {Kitamoto}, {Iga}, {Negoro}, \&
  {Terada}}]{miy92}
{Miyamoto}, S., {Kitamoto}, S., {Iga}, S., {Negoro}, H., \& {Terada}, K. 1992,
  \apjl, 391, L21

\bibitem[{{Oosterbroek} {et~al.}(1997){Oosterbroek}, {van der Klis}, {van
  Paradijs}, {Vaughan}, {Rutledge}, {Lewin}, {Tanaka}, {Nagase}, {Dotani},
  {Mitsuda}, \& {Miyamoto}}]{oost97}
{Oosterbroek}, T., {van der Klis}, M., {van Paradijs}, J., {et~al.} 1997, \aap,
  321, 776

\bibitem[{{Osborne} {et~al.}(2001){Osborne}, {Borozdin}, {Trudolyubov},
  {Priedhorsky}, {Soria}, {Shirey}, {Hayter}, {La Palombara}, {Mason},
  {Molendi}, {Paerels}, {Pietsch}, {Read}, {Tiengo}, {Watson}, \&
  {West}}]{osb01}
{Osborne}, J.~P., {Borozdin}, K.~N., {Trudolyubov}, S.~P., {et~al.} 2001, \aap,
  378, 800

\bibitem[{{Shahbaz} {et~al.}(1994){Shahbaz}, {Ringwald}, {Bunn}, {Naylor},
  {Charles}, \& {Casares}}]{shab94}
{Shahbaz}, T., {Ringwald}, F.~A., {Bunn}, J.~C., {et~al.} 1994, \mnras, 271,
  L10

\bibitem[{{{\v S}imon}(2003)}]{sim03}
{{\v S}imon}, V. 2003, \aap, 405, 199

\bibitem[{{Stella} {et~al.}(1987){Stella}, {Priedhorsky}, \& {White}}]{stel87}
{Stella}, L., {Priedhorsky}, W., \& {White}, N.~E. 1987, \apjl, 312, L17

\bibitem[{{Str{\" u}der} {et~al.}(2001){Str{\" u}der}, {Briel}, {Dennerl},
  {Hartmann}, {Kendziorra}, {Meidinger}, {Pfeffermann}, {Reppin}, {Aschenbach},
  {Bornemann}, {Br{\" a}uninger}, {Burkert}, {Elender}, {Freyberg}, {Haberl},
  {Hartner}, {Heuschmann}, {Hippmann}, {Kastelic}, {Kemmer}, {Kettenring},
  {Kink}, {Krause}, {M{\" u}ller}, {Oppitz}, {Pietsch}, {Popp}, {Predehl},
  {Read}, {Stephan}, {St{\" o}tter}, {Tr{\" u}mper}, {Holl}, {Kemmer},
  {Soltau}, {St{\" o}tter}, {Weber}, {Weichert}, {von Zanthier},
  {Carathanassis}, {Lutz}, {Richter}, {Solc}, {B{\" o}ttcher}, {Kuster},
  {Staubert}, {Abbey}, {Holland}, {Turner}, {Balasini}, {Bignami}, {La
  Palombara}, {Villa}, {Buttler}, {Gianini}, {Lain{\' e}}, {Lumb}, \&
  {Dhez}}]{stru01}
{Str{\" u}der}, L., {Briel}, U., {Dennerl}, K., {et~al.} 2001, \aap, 365, L18

\bibitem[{{Tomsick} {et~al.}(2003){Tomsick}, {Corbel}, {Fender}, {Miller},
  {Orosz}, {Rupen}, {Tzioumis}, {Wijnands}, \& {Kaaret}}]{tom03}
{Tomsick}, J.~A., {Corbel}, S., {Fender}, R., {et~al.} 2003, \apjl, 597, L133

\bibitem[{{Trudolyubov} {et~al.}(2002){Trudolyubov}, {Borozdin}, {Priedhorky},
  {Osborne}, {Mason}, \& {Cordova}}]{tru02}
{Trudolyubov}, S., {Borozdin}, K.~N., {Priedhorky}, W.~C., {et~al.} 2002, ApJL,
  581, L27

\bibitem[{{Turner} {et~al.}(2001){Turner}, {Abbey}, {Arnaud}, {Balasini},
  {Barbera}, {Belsole}, {Bennie}, {Bernard}, {Bignami}, {Boer}, {Briel},
  {Butler}, {Cara}, {Chabaud}, {Cole}, {Collura}, {Conte}, {Cros}, {Denby},
  {Dhez}, {Di Coco}, {Dowson}, {Ferrando}, {Ghizzardi}, {Gianotti}, {Goodall},
  {Gretton}, {Griffiths}, {Hainaut}, {Hochedez}, {Holland}, {Jourdain},
  {Kendziorra}, {Lagostina}, {Laine}, {La Palombara}, {Lortholary}, {Lumb},
  {Marty}, {Molendi}, {Pigot}, {Poindron}, {Pounds}, {Reeves}, {Reppin},
  {Rothenflug}, {Salvetat}, {Sauvageot}, {Schmitt}, {Sembay}, {Short},
  {Spragg}, {Stephen}, {Str{\" u}der}, {Tiengo}, {Trifoglio}, {Tr{\" u}mper},
  {Vercellone}, {Vigroux}, {Villa}, {Ward}, {Whitehead}, \& {Zonca}}]{turn01}
{Turner}, M.~J.~L., {Abbey}, A., {Arnaud}, M., {et~al.} 2001, \aap, 365, L27

\bibitem[{{Uttley} {et~al.}(2002){Uttley}, {McHardy}, \& {Papadakis}}]{utt02}
{Uttley}, P., {McHardy}, I.~M., \& {Papadakis}, I.~E. 2002, \mnras, 332, 231



\bibitem[{{van den Bergh}(2000)}]{vdb00}
{van den Bergh}, S. 2000, The galaxies of the Local Group,  Cambridge University Press, Cambridge
  Astrophysics Series Series, vol no: 35

\bibitem[{{Van der Klis}(1994)}]{vdk94}
{van der Klis}, M. 1994, \apjs, 92, 511

\bibitem[{{van Paradijs} \& {McClintock}(1995)}]{vm95}
{van Paradijs}, J. \& {McClintock}, J.~E. 1995, {X-ray Binaries} ({Cambridge
  University Press}), 58--125

\bibitem[{{Williams} {et~al.}(2003){Williams}, {Garcia}, {Kong}, {Primini},
  {King}, {Di Stefano}, \& {Murray}}]{will03}
{Williams}, B.~F., {Garcia}, M.~R., {Kong}, A.~K.~H., {et~al.} 2003, ApJ in
  press, astro-ph/0306421 v2

\bibitem[{{Wilson} {et~al.}(2003){Wilson}, {Patel}, Kouveliotou, G., {van der
  Klis}, {Lewin}, Belloni, \& Mendez}]{wil03}
{Wilson}, C.~A., {Patel}, S.~K., Kouveliotou, C., {et~al.} 2003, \apj, 596,
  1220

\bibitem[{{Zdziarski} {et~al.}(2004){Zdziarski}, {Gierlinski}, {Mikolajewska},
  {Wardzinski}, {Smith}, {Harmon}, \& {Kitamoto}}]{zdz04}
{Zdziarski}, A.~A., {Gierlinski}, M., {Mikolajewska}, J., {et~al.} 2004, astro-ph/0402380

\end{thebibliography}
\end{document}